\documentclass[12pt,a4paper]{article}

\usepackage[utf8]{inputenc}
\usepackage[T1]{fontenc}
\usepackage{lmodern}
\usepackage[english]{babel}

\usepackage{graphicx}
\usepackage{amsmath}
\usepackage{amssymb}
\usepackage{textcomp}
\usepackage{gensymb}          
\usepackage{siunitx}
\usepackage{booktabs}
\usepackage{array}
\usepackage{caption}
\usepackage{authblk}

\newif\ifblind\blindfalse
\ifdefined\BLIND\blindtrue\fi
\usepackage[margin=2.5cm]{geometry}
\usepackage{setspace}
\usepackage[hidelinks]{hyperref}

\graphicspath{{figs/}}
\title{\textbf{Effect of Mangosteen Peel
(\textit{Garcinia mangostana} L.) and Grass Jelly Leaf
(\textit{Cyclea barbata} L. Miers) Natural Dye Mixing Ratio on the
Optical, Morphological, and Photovoltaic Properties of
Dye-Sensitized Solar Cells (DSSC)}}

\ifblind
\author{}
\else
\author[1,6,*]{Eka Nurfani}
\author[1]{Haifa Salsabila}
\author[1]{Dwiky I. Bakhtiar}
\author[1]{Wahyu S. Sipahutar}
\author[2]{M. Alvien Ghifari}
\author[3]{Rishal Asri}
\author[4]{Meqorry Yusfi}
\author[5,6]{Tarmizi Taher}
\author[1,6]{Aditya Rianjanu}
\author[7]{Robi Kurniawan}

\affil[1]{Department of Materials Engineering, Faculty of Industrial Technology, Institut Teknologi Sumatera, South Lampung 35365, Indonesia}
\affil[2]{Department of Chemistry, Faculty of Sciences, Institut Teknologi Sumatera, South Lampung 35365, Indonesia}
\affil[3]{Department of Energy System Engineering, Faculty of Industrial Technology, Institut Teknologi Sumatera, South Lampung 35365, Indonesia}
\affil[4]{Department of Physics, Faculty of Mathematics and Natural Sciences, Universitas Andalas, Padang 25175, Indonesia}
\affil[5]{Department of Environmental Engineering, Faculty of Infrastructure and Regional Technology, Institut Teknologi Sumatera, South Lampung 35365, Indonesia}
\affil[6]{Center for Green and Sustainable Materials, Institut Teknologi Sumatera, Terusan Ryacudu, Way Huwi, Jati Agung, Lampung Selatan 35365, Indonesia}
\affil[7]{Department of Physics, Faculty of Mathematics and Natural Sciences, Universitas Negeri Malang, Malang 65145, Indonesia}
\affil[*]{Corresponding author. E-mail: \href{mailto:eka.nurfani@mt.itera.ac.id}{eka.nurfani@mt.itera.ac.id} (EN)}
\fi

\date{}

\begin{document}
\maketitle

\newpage
\doublespacing

\section*{Abstract}
Dye-sensitized solar cells (DSSCs) are promising low-cost and
environmentally friendly photovoltaic devices. This study explores the
effect of the mixing ratio of natural dyes extracted from mangosteen peel
(MP) and grass jelly leaves (GJL) on the optical, chemical,
morphological, and photovoltaic properties of DSSCs. Dye solutions were
prepared from crude extracts at a fixed concentration of
0.6~g\,mL\textsuperscript{$-1$}, with MP:GJL volume ratios varied from
3:1 to 1:3. UV-Vis absorption spectra showed that MP contributes to the
UV--blue region (350--450~nm) and GJL to the red region (600--700~nm),
giving the mixtures a broader, more panchromatic absorption.
Fourier-transform infrared (FTIR) spectra confirmed that both extracts
share hydroxyl (--OH) and carbonyl/carboxyl (C=O, C--O) groups capable of
anchoring the pigments onto TiO\textsubscript{2}, with a stronger
glycosidic C--O--C band for MP. Optical and FESEM imaging revealed that
balanced and GJL-rich mixtures form more uniform, crack-free films than
single dyes. Current--voltage ($J$--$V$) measurements of three
independently fabricated cells per composition ($n=3$) under identical
white-light illumination showed that the 1:3 (MP:GJL) ratio gave the
highest apparent power conversion efficiency, about 1.3 and 1.7 times
that of the single MP and GJL dyes, respectively. Because the input
power was estimated photometrically (lux) rather than with a calibrated
AM1.5G source, these efficiencies are apparent (relative) indicators for
comparing samples, not absolute values. The improvement is attributed to
the complementary absorption of the two pigments, while direct
electrochemical confirmation of the charge-transfer mechanism remains
for future work.

\vspace{0.5em}
\noindent\textbf{Keywords:} DSSC, natural dye, mangosteen peel, grass
jelly leaves

\section{Introduction}

The rapid increase in global energy consumption, driven by industrial
growth and population expansion, has led to a significant reliance on
fossil fuels. However, fossil energy sources are non-renewable,
environmentally harmful, and increasingly scarce, necessitating the
exploration of sustainable alternatives~\cite{ref1}. Solar energy has
emerged as one of the most abundant and promising renewable resources,
potentially supplying more energy than the world's total
demand~\cite{ref2}. Various photovoltaic technologies, including
crystalline silicon~\cite{ref3}, thin-film solar cells~\cite{ref4}, and
emerging third-generation devices~\cite{ref5}, have been developed to
harness solar power. Among them, dye-sensitized solar cells (DSSC),
first reported by O'Regan and Gr\"atzel in 1991, have gained attention
due to their relatively simple fabrication process, low production
costs, and the possibility of using eco-friendly materials compared to
conventional silicon-based solar cells~\cite{ref6}.

One of the key components of DSSC is the dye sensitizer, which plays a
crucial role in light absorption and electron excitation. While
ruthenium-based synthetic dyes have demonstrated high efficiencies,
their high cost, scarcity, and toxicity limit large-scale
application~\cite{ref7}. In addition to metal-complex dyes, synthetic
organic dyes such as naphthalimide-based sensitizers have also been
investigated due to their tunable molecular structures and strong light
absorption properties~\cite{ref8}. However, these dyes often require
complex synthesis procedures and may suffer from stability issues.
Consequently, natural dyes derived from plant pigments have been widely
explored as low-cost and sustainable alternatives~\cite{ref9}. Recent
studies have shown that anthocyanins~\cite{ref10,ref11,ref12},
betalains~\cite{ref13,ref14,ref15,ref16}, and
chlorophyll~\cite{ref17,ref18} are effective sensitizers due to their
strong absorption in the visible region and their ability to transfer
electrons to TiO\textsubscript{2} photoanodes. Mangosteen peel
(\textit{Garcinia mangostana} L.) is rich in
anthocyanins~\cite{ref19,ref20,ref21}, which provide excellent photon
absorption and redox properties, while grass jelly leaves
(\textit{Cyclea barbata} L. Miers) contain high levels of
chlorophyll~\cite{ref22}. Several recent works have reported enhanced
DSSC performance by co-sensitization, i.e., combining multiple pigments
to broaden the absorption range and improve device
efficiency~\cite{ref17,ref38,ref39,ref40}. For example, chlorophyll and
anthocyanin have been co-sensitized to extend spectral
coverage~\cite{ref17}, and fruit- and flower-derived dyes have been
combined to raise the photocurrent of natural-dye
DSSCs~\cite{ref38,ref39,ref40}.

Despite these advances, most co-sensitization studies have focused on
pigment pairs and plant sources other than the mangosteen-peel/grass-jelly-leaf
combination, and few have systematically optimized the mixing ratio of
an anthocyanin-rich and a chlorophyll-rich extract. In this work,
natural dyes extracted from mangosteen peel (MP) and grass jelly leaves (GJL) were
systematically combined in various volume ratios (3:1, 2:1, 1:1, 1:2,
and 1:3, together with the single-dye controls 3:0 and 0:3) to
investigate their optical absorption, functional groups, dye adsorption
morphology, and photovoltaic response. The novelty of this study lies
in (i) pairing the anthocyanin/$\alpha$-mangostin-rich mangosteen peel
with chlorophyll-rich \textit{Cyclea barbata} grass jelly leaf, a
combination not previously reported for co-sensitized DSSCs, and (ii)
mapping how the anthocyanin-to-chlorophyll balance governs dye
deposition on TiO\textsubscript{2} and the resulting light-harvesting
behavior. By identifying the dye ratio that maximizes the apparent
photovoltaic response, this research contributes to developing low-cost,
eco-friendly, and sustainable solar energy conversion technologies based
on locally available agricultural by-products. We emphasize at the
outset that, because the illumination was measured photometrically
rather than with a calibrated AM1.5G simulator, all efficiencies in this
work are reported as apparent (relative) values used only to rank the
compositions tested under identical conditions.

\section{Materials and Methods}

\subsection{Substrate Preparation}

Figure~\ref{fig:1} is a schematic experimental design to fabricate the
DSSC device, including (1) TiO\textsubscript{2} film, (2) dyes with
various MP:GJL ratios, (3) electrolyte layer, and (4) graphite as back
electrode. Firstly, indium tin oxide (ITO)-coated glass (2.0$\times$1.5~cm) was prepared
and subjected to a rigorous three-stage ultrasonic cleaning process to
remove organic, ionic, and particulate contaminants. The cleaning
sequence began with an acetone treatment for 10 minutes to dissolve
organic impurities, followed by an ethanol rinse for another 10 minutes
to remove residual contaminants and degrease the surface, and concluded
with a final wash in deionized water for 10 minutes to eliminate any
remaining ionic species or solvent traces~\cite{ref23,ref24,ref25}.
Subsequently, the cleaned ITO substrates were thoroughly dried on a hot
plate at 100\,\celsius{} for one hour to ensure complete moisture
evaporation before further use. The substrates were handled exclusively
with tweezers throughout the process to prevent surface contamination.

\subsection{Fabrication of TiO\textsubscript{2} Film}

\begin{figure}[htbp]
\centering
\includegraphics[width=0.70\linewidth]{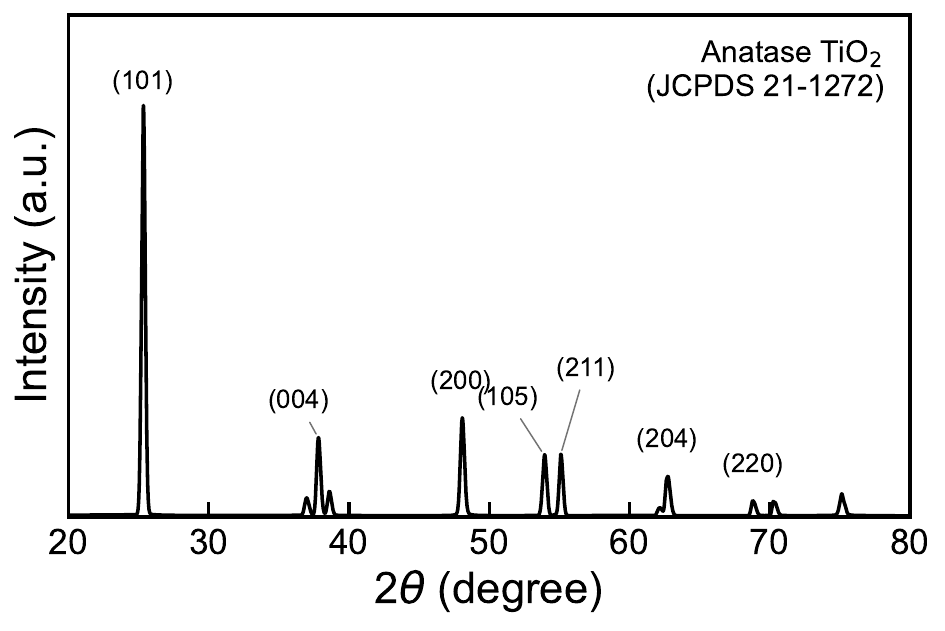}
\caption{XRD pattern of the commercial TiO\textsubscript{2} powder used
as the photoanode material. All diffraction peaks are indexed to the
anatase phase (JCPDS card No. 21-1272); no rutile or brookite peaks are
observed.}
\label{fig:xrd}
\end{figure}

Commercial titanium dioxide (TiO\textsubscript{2}) anatase powder (pro
analysis grade, Sigma-Aldrich) was used as the semiconductor material
for the photoanode preparation. The same commercial powder has been
characterized previously, with a primary particle size of approximately
130~nm~\cite{ref26}. The crystalline phase of the
TiO\textsubscript{2} was further confirmed by X-ray diffraction (XRD)
analysis, as shown in Figure~\ref{fig:xrd}. The diffraction peaks
observed at
$2\theta \approx 25.3\degree, 37.8\degree, 48.0\degree, 53.9\degree,
55.1\degree, 62.7\degree$, and $68.8\degree$ correspond to the (101),
(004), (200), (105), (211), (204), and (220) planes of anatase
TiO\textsubscript{2}, respectively, which are consistent with the
standard Joint Committee on Powder Diffraction Standards (JCPDS) card No. 21-1272. No diffraction peaks corresponding to
rutile or brookite phases were detected, indicating that the
TiO\textsubscript{2} used in this study is predominantly anatase.
TiO\textsubscript{2} was chosen as a semiconductor material because it
allows interaction with redox electrolytes. This material has a wide
band gap value, where semiconductors also play a role in the electron
transfer process, and as a place to attach the color layer that will
be excited due to exposure to light. In making this
TiO\textsubscript{2} paste, 5 grams of TiO\textsubscript{2} are mixed
with 25~mL of ethanol as a solvent, then stirred on a magnetic stirrer
at 600~rpm for 2 hours at room temperature~\cite{ref26}; the paste
viscosity was not quantitatively measured. The cleaned
ITO substrate was then coated with TiO\textsubscript{2} paste using a
single spin-coating cycle at a speed of 3000~rpm for 15 seconds, which
was then heated on a hot plate for 1 hour at a temperature of
300\,\celsius{}. No compact TiO\textsubscript{2} blocking layer was
deposited between the ITO and the mesoporous TiO\textsubscript{2} layer.
Cross-sectional field-emission scanning electron microscopy (FESEM; Section~\ref{sec:morph}) gave a
TiO\textsubscript{2}/dye film thickness of approximately
20~\textmu m, which is dominated by the mesoporous
TiO\textsubscript{2} layer; the adsorbed dye contributes a negligible
fraction of this value. Because the stack thickness is governed by the
TiO\textsubscript{2} deposition, which was identical for all samples,
it was not resolved individually for every MP:GJL ratio and is reported
as a common value.

\subsection{Dye Preparation}

The first step in the dye extraction stage was washing the mangosteen
peels and grass jelly leaves to remove any impurities
adhering to their surfaces. The next step was drying, which was carried
out in an oven for 6 hours at 100\,\celsius{}. The samples were ground
into fine powder using a blender until a homogeneous powder was
obtained. The next step was weighing the powder. Eighteen grams of
mangosteen peel and grass jelly leaves powder were measured and
dissolved in 30~mL of ethanol in a beaker, which was then tightly
sealed with aluminum foil. The beaker was placed on a magnetic stirrer
to facilitate mixing and extracting the dye content from the samples.
This process was conducted for 2 hours at 70\,\celsius{} with a
stirring speed of 600~rpm. Because anthocyanins and chlorophyll are
sensitive to heat, light, oxygen, and pH~\cite{ref19,ref35}, the
extraction was performed in aluminum-foil--sealed vessels and the
extracts were stored in the dark and used promptly; however, the
extraction and drying temperatures (70\,\celsius{} and 100\,\celsius{},
respectively) were not further optimized to maximize pigment retention,
and some degradation of thermolabile pigments cannot be excluded. The
homogeneous solution was filtered using filter paper with a diameter of
25~\textmu m to separate the residue from the dye extract solution.
After obtaining a clear dye extract free from residues, mixtures of the
extracts were prepared with different ratios of mangosteen peel and
grass jelly leaves dye. The volume ratios of MP:GJL were 3:1, 2:1, 1:1,
1:2, and 1:3. The resulting dye solutions were used as crude extracts
with an approximate concentration of 0.6~g\,mL\textsuperscript{$-1$}, representing the mass of
plant extract per unit volume of solvent. It should be noted that this
concentration does not correspond to the exact concentration of active
chromophores such as anthocyanins, $\alpha$-mangostin, or chlorophyll,
but rather reflects the total extract content; the individual
chromophore concentrations were not quantified in this study, so the
mixing ratios correspond to \emph{relative} extract compositions rather
than compositions equalized by chromophore content. To infuse the dye
layer, the glass was soaked in 3~mL dye solution for 24 hours to obtain
Dye/TiO\textsubscript{2}/ITO.

\subsection{Electrolyte Preparation}

The electrolyte was prepared using iodide, specifically 1.66~g of
potassium iodide (KI), mixed with 20~mL of ethylene glycol. The mixture
was stirred using a magnetic stirrer for 2 hours at a speed of
300~rpm. During the stirring process, 0.252~g of iodine
(I\textsubscript{2}) was added and stirred until a homogeneous solution
was achieved, according to the previous study~\cite{ref27}. The
selected chemicals were used for the following reasons: KI provides
$\mathrm{I^{-}}$ ions and iodine provides $\mathrm{I_{2}}$ molecules,
which together form the essential $\mathrm{I^{-}/I_{3}^{-}}$ redox
couple for electron circulation via the reaction
$\mathrm{I_{2} + I^{-} \rightleftharpoons I_{3}^{-}}$. Ethylene glycol,
a polar solvent, was used to maintain the stability and solubility of
the electrolyte system efficiently.

\subsection{Preparation of Graphite Layer}

The graphite carbon paste was prepared by mixing 1.88~g of graphite,
0.56~g of polyvinylidene fluoride (PVDF) as a binder, and 9.4~mL of
dimethylformamide (DMF) as a solvent~\cite{ref28,ref29}. The mixture
was stirred using a magnetic stirrer on a hot plate at 80\,\celsius{}
for 30 minutes to ensure initial dissolution and homogeneity.
Subsequently, a second stirring step was conducted at a higher
temperature of 200\,\celsius{} for another 30 minutes at a speed of
600~rpm. The resulting homogeneous carbon paste was then deposited onto
the conductive surface of an ITO-coated glass substrate. The deposition
was performed using a spin coater at a rotational speed of 2000~rpm for
15 seconds. This process ensures the formation of a thin, uniform
catalytic counter electrode layer essential for the DSSC assembly. The
graphite layer was then heated on the hotplate at 300\,\celsius{} for
one hour to obtain Graphite/ITO.

\subsection{DSSC Assembly}

The dye-adsorbed semiconductor substrate (Dye/TiO\textsubscript{2}/ITO)
was carefully assembled into a complete cell by placing it
face-to-face with the carbon-coated counter electrode (Graphite/ITO).
The two substrates were offset slightly to create an overlapping active
area and securely clipped together on both sides with binder clips,
forming a stable sandwich-type configuration. No dedicated spacer or
sealant was used; the cell was an open, clip-assembled device, so the
electrode gap was defined by the film thickness and the clip pressure
rather than by a fixed spacer. The active area (1.2~cm\textsuperscript{2})
was set by the geometric overlap between the dye-coated
TiO\textsubscript{2} region and the graphite counter electrode, and was
further delimited during the measurement by an aperture mask of the same
area placed over the illuminated face, so that the area used to compute
$J_{sc}$ corresponds to the region actually under illumination. Finally,
a single drop of approximately 50~\textmu L of the prepared redox
electrolyte solution was delivered by micropipette to the edge of the
assembled cell and drawn into the gap between the
substrates via capillary action, ensuring uniform
distribution between the semiconductor and carbon layers. Excess
electrolyte was expelled at the edges as the binder clips were
tightened, so the volume retained within the
$\approx$20~\textmu m inter-electrode gap over the
1.2~cm\textsuperscript{2} active area is of the order of
2--5~\textmu L. Once the
electrolyte thoroughly permeated the internal space, the assembled DSSC
was immediately ready for electrical characterization and performance
measurement.

\subsection{Characterization Methods}
\label{sec:charac}

After extracting the dyes for use in the DSSC, the dye solutions were
diluted 30 times from their original concentration. This dilution was
necessary to ensure the absorbance peaks of the dyes fell within the
detectable range of the instrument. Undiluted dyes with excessively
high concentrations would result in truncated absorbance peaks, leading
to inaccurate measurements of maximum absorption for comparison. The
diluted dye solutions were subsequently analyzed using a
ultraviolet-visible spectrophotometer (Shimadzu UV1280) to determine
their maximum absorption wavelength and corresponding absorbance
values. The functional groups of the MP and GJL dye extracts were
identified by Fourier-transform infrared (FTIR) spectroscopy
(Shimadzu IRSpirit) over the 400--4000~cm\textsuperscript{$-1$} range. Moreover,
morphological and cross-sectional analysis of
Dye/TiO\textsubscript{2}/ITO were performed using an optical microscope
(Amscope) and FESEM (Thermo Scientific Quattro S).

Current--voltage ($J$--$V$) measurements were carried out using a
Source Measurement Unit (SMU, Keithley 2450) over a defined active area
of $A = 1.2$~cm\textsuperscript{2}. For each MP:GJL composition, three
independently fabricated cells were measured ($n=3$), and the
photovoltaic parameters are reported as mean $\pm$ standard deviation
(SD). The light source was a 4000~K full-spectrum white-LED lamp
(Sansi PAR20, E27, 10~W, 1000~lm), whose illuminance at the device
surface was monitored with a digital luxmeter and maintained at
approximately $L = 2.5\times10^{5}$~lux throughout the
measurements. Because a calibrated AM1.5G solar simulator
(100~mW\,cm\textsuperscript{$-2$}) was not available, the incident
irradiance $E$ was estimated from the measured illuminance using the
luminous efficacy of radiation of this lamp
($K \approx 256$~lm\,W\textsuperscript{$-1$}, derived from its rated
luminous flux and photon flux), following the lux-to-irradiance
conversion procedure of Michael et al.~\cite{ref41} on the same
white-LED measurement bench described in~\cite{ref42}:

\begin{equation}
E = \frac{L}{K} , \qquad P_{in} = E \times A ,
\label{eq:pin}
\end{equation}

\noindent where $A$ is the active area. For
$L = 2.5\times10^{5}$~lux, $K = 256$~lm\,W\textsuperscript{$-1$} and
$A = 1.2$~cm\textsuperscript{2}, this yields
$E \approx 98$~mW\,cm\textsuperscript{$-2$} ($\approx 0.98$~sun) and
$P_{in} \approx 0.12$~W over the active area. Two caveats must be
stressed. First, although $E$ happens to lie close to the AM1.5G
reference irradiance of 100~mW\,cm\textsuperscript{$-2$}, the
\emph{spectrum} of a 4000~K white LED differs substantially from
AM1.5G: it carries essentially no near-infrared component, so the
red/near-infrared absorption band of chlorophyll is under-excited
relative to standard conditions, whereas the blue-green region relevant
to anthocyanin is comparatively well covered. Second, because
lux is a
photometric quantity weighted by the human photopic response and
strongly dependent on the spectral distribution of the lamp, this
conversion is only an approximation and was not verified against a
calibrated reference cell. Consequently, the efficiency values
reported in this work are \emph{apparent} (relative) values intended to
reflect performance trends among samples measured under identical
conditions, and they should not be compared directly with AM1.5G-calibrated
efficiencies. All testing conditions (lamp, geometry, active area,
and measurement sequence) were kept constant across all samples to
ensure the validity of the relative comparison.

The key parameters evaluated in the DSSC performance included the
open-circuit voltage ($V_{oc}$), short-circuit current ($I_{sc}$),
short-circuit current density ($J_{sc} = I_{sc}/A$), fill factor
(\textit{FF}), and apparent power conversion efficiency ($\eta$),
calculated as

\begin{equation}
\eta = \frac{V_{oc} \, I_{sc} \, FF}{P_{in}} \times 100\% .
\label{eq:eta}
\end{equation}

As a worked example, the best-performing 1:3 (MP:GJL) cell had
$V_{oc}=0.18$~V, $I_{sc}=0.18$~mA (hence
$J_{sc}=I_{sc}/A = 0.18~\mathrm{mA}/1.2~\mathrm{cm^2} = 0.15$~mA\,cm\textsuperscript{$-2$}),
and $FF=0.28$. Substituting these values together with
$P_{in} \approx 0.12$~W from Eq.~\eqref{eq:pin} into Eq.~\eqref{eq:eta}
gives

\begin{equation*}
\eta = \frac{0.18~\mathrm{V} \times 1.8\times10^{-4}~\mathrm{A} \times 0.28}
{0.12~\mathrm{W}} \times 100\%
= \frac{9.1\times10^{-6}~\mathrm{W}}{0.12~\mathrm{W}} \times 100\%
\approx 7.7\times10^{-3}\% ,
\end{equation*}

\noindent which is the apparent efficiency of the best single 1:3 cell.
The value tabulated for the 1:3 composition in Table~\ref{tab:1}
($7.47\times10^{-3}\%$) is the mean over the three independent cells of
that composition and is therefore slightly lower. These values are
internally consistent relative figures of merit and are not comparable
with AM1.5G-calibrated efficiencies.

\begin{figure}[htbp]
\centering
\includegraphics[width=0.92\linewidth]{fig-1-schematic.pdf}
\caption{Schematic experiment to synthesize all layers in DSSC,
including (1) mesoporous TiO\textsubscript{2}, (2) dye sensitizer, (3)
electrolyte, and (4) graphite-based counter electrode.}
\label{fig:1}
\end{figure}

\section{Results and Discussion}

\subsection{Absorption Properties of Dye}

\begin{figure}[htbp]
\centering
\includegraphics[width=0.62\linewidth]{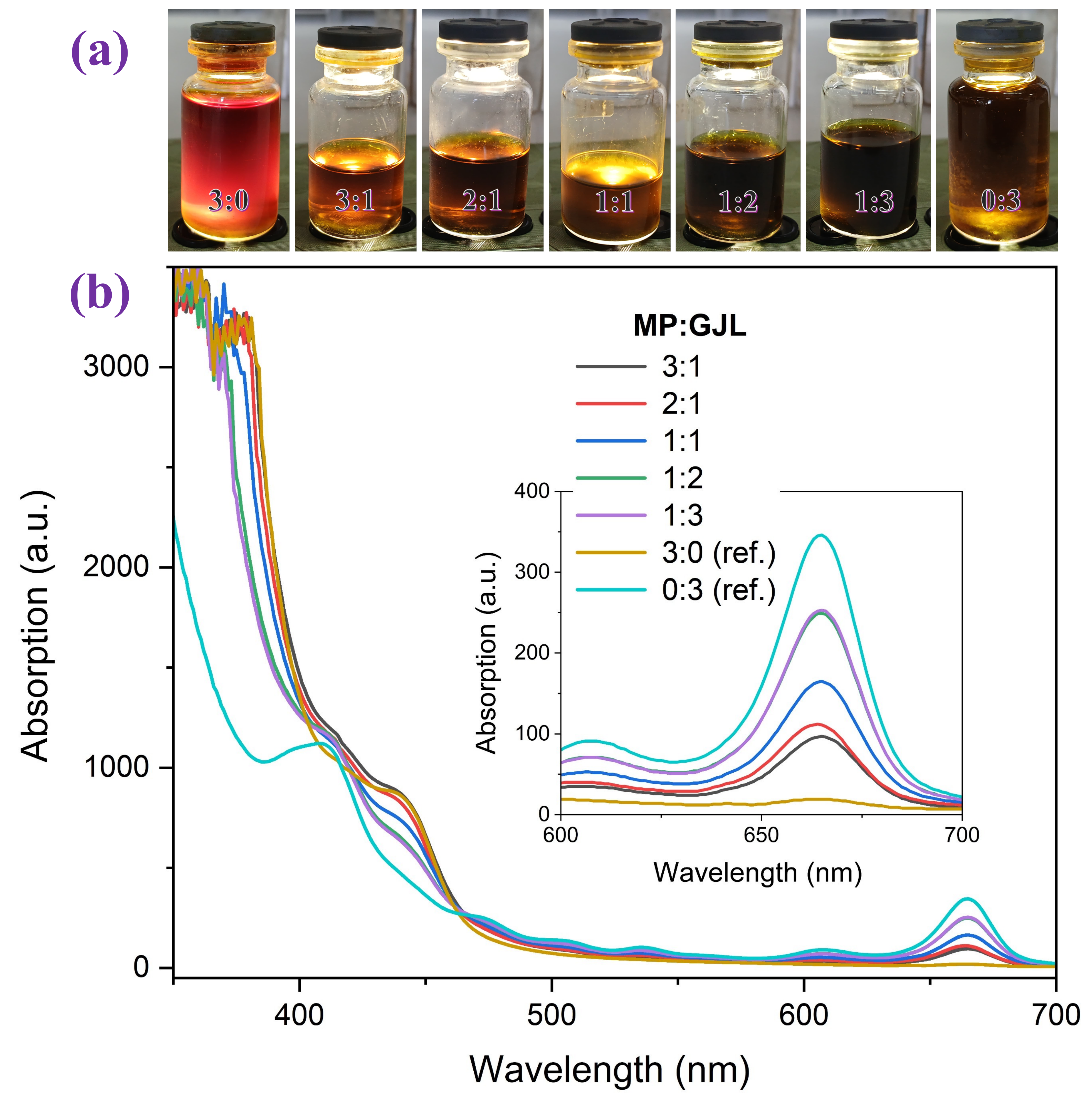}
\caption{(a) Mixed dye with the MP:GJL ratio of 3:0, 3:1, 2:1, 1:1,
1:2, 1:3, and 0:3. (b) Absorption spectra of all mixed dyes in the
350--700~nm range. Inset of (b) is the zoom view for absorption spectra
in the 600--700~nm range.}
\label{fig:2}
\end{figure}

Figure~\ref{fig:2} presents the color of MP and GJL dyes, their
mixture, and UV-Vis absorption spectra of the pigments at various
volume ratios. Figure~\ref{fig:2}a shows the visual appearance of the
mixed dye solutions prepared with different volume ratios of MP to GJL
(3:0, 3:1, 2:1, 1:1, 1:2, 1:3, and 0:3). The photographs clearly
indicate a distinct variation in color as the mixing ratio changes,
reflecting differences in the light absorption characteristics of the
dyes. The MP-rich solutions (3:0 and 3:1) exhibit an intense reddish
coloration. At the same time, the color gradually shifts to
orange-yellow in the intermediate mixtures (2:1 and 1:1). Further
increase of the GJL content (1:2, 1:3, and 0:3) results in a darker
yellow-brown hue. This progressive color change implies that the
spectral properties of the mixed dyes can be tuned by adjusting the
MP:GJL ratio, suggesting an improvement in their absorption behavior.

The absorption spectra show the complete absorption profiles of the
individual and composite dyes (Figure~\ref{fig:2}b). The spectrum for
the pure MP and GJL dye exhibits a characteristic absorption band
attributed to chlorophyll (610 and 665~nm)~\cite{ref30}, anthocyanin
(440--460~nm and 530--550~nm)~\cite{ref31}, and $\alpha$-mangostin
(350--370~nm)~\cite{ref20,ref21,ref31}. Pure MP dye (3:0) shows the
strongest absorption related to anthocyanin and $\alpha$-mangostin
compared to GJL. There is no chlorophyll peak observed. Conversely, the
spectrum for the pure GJL dye (0:3) displays a prominent absorption
peak related to chlorophyll. The spectra of the blended dyes
consistently show two distinct absorption features corresponding to
both parental dyes. This indicates that the mixing process successfully
combines the optical properties of both extracts without significant
chemical interaction that would alter their fundamental absorption
characteristics. As the volume ratio of GJL to MP increases (from 3:1
to 1:3), a clear trend is observed: the intensity of the
chlorophyll-related peak in the red region increases progressively,
while the intensity of the anthocyanin peak in the blue region
decreases accordingly.

The inset of Figure~\ref{fig:2}b, a magnified view of the
600--700~nm region, provides a closer examination of these
changes in absorption intensity. This detailed view illustrates the
systematic enhancement of the red absorption band with higher GJL
content in the mixtures. The complementary absorption profiles of MP
and GJL are key findings. The blending strategy effectively broadens
the overall light-harvesting window across the visible spectrum. This
strategy creates a more panchromatic sensitizer, a fundamental
requirement for enhancing the photon capture and the photocurrent in
DSSCs. The observed optical properties
provide a plausible rationale for the improved apparent photovoltaic
performance discussed below, particularly for the 1:3 (MP:GJL) ratio,
which achieved the highest apparent power conversion efficiency.

\subsection{Functional Group Analysis (FTIR)}

\begin{figure}[htbp]
\centering
\includegraphics[width=0.70\linewidth]{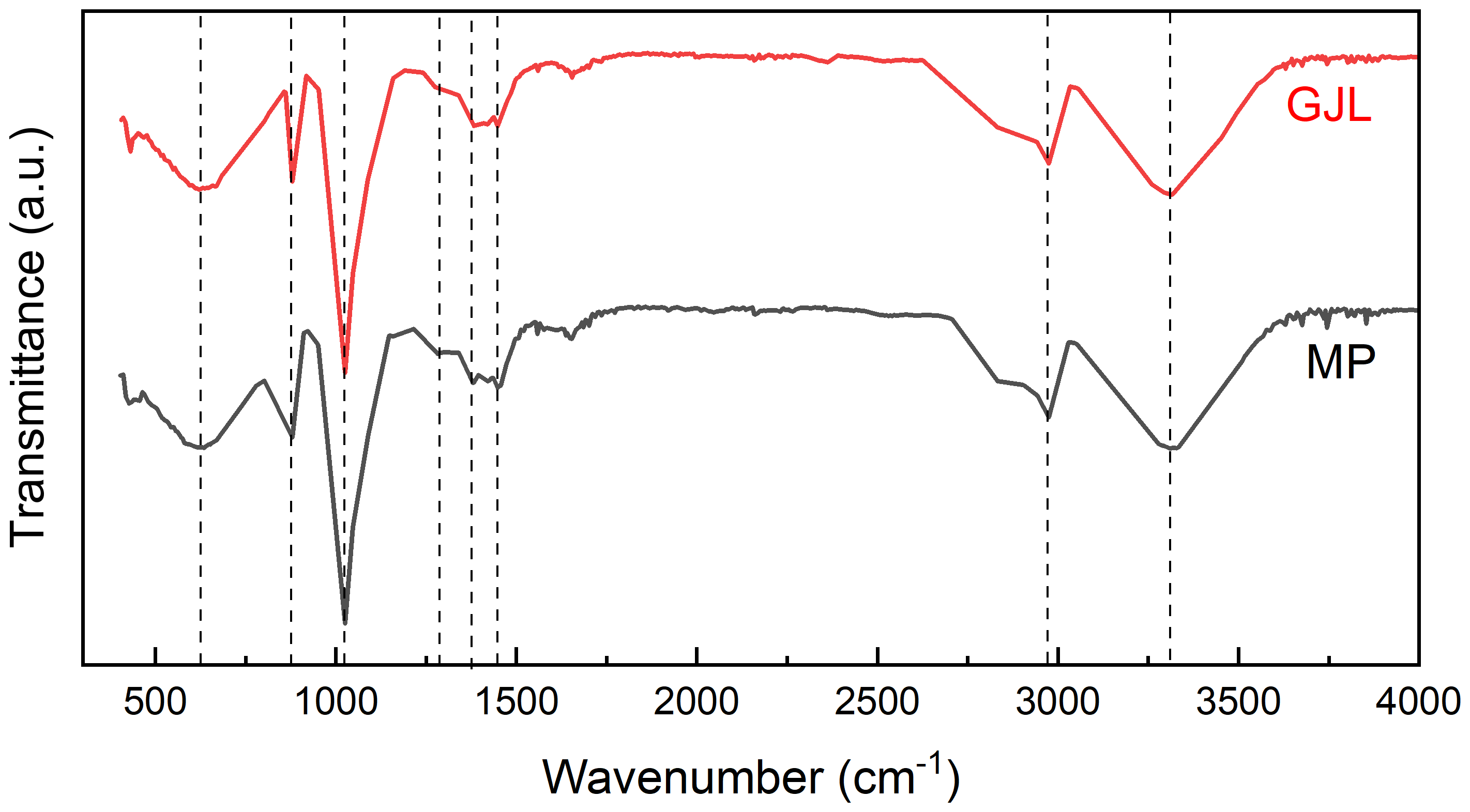}
\caption{FTIR transmittance spectra of the pure mangosteen peel (MP) and
grass jelly leaves (GJL) natural dye extracts. Dashed lines mark the
common absorption bands.}
\label{fig:ftir}
\end{figure}

Figure~\ref{fig:ftir} shows the FTIR transmittance spectra of the pure
MP and GJL dye extracts over the
400--4000~cm\textsuperscript{$-1$} range. Both extracts exhibit a
similar set of vibrational bands, indicating that they share common
oxygen-rich functional groups characteristic of plant-derived phenolic
and polysaccharide constituents. A broad band centered near
$\sim$3300~cm\textsuperscript{$-1$} is assigned to O--H stretching of
hydroxyl groups from phenolics, carboxylic acids, and adsorbed
water~\cite{ref35,ref37}, while the band around
$\sim$2920~cm\textsuperscript{$-1$} corresponds to aliphatic C--H
stretching~\cite{ref35,ref37}. The features in the
$\sim$1540--1600~cm\textsuperscript{$-1$} region are attributed to
aromatic C=C and conjugated C=O vibrations of the chromophore rings---the
flavylium/aromatic core of anthocyanin and the xanthone of
$\alpha$-mangostin in MP, and the porphyrin/carbonyl moieties of
chlorophyll in GJL~\cite{ref35,ref36,ref37}. Bands near
$\sim$1380, $\sim$1270, and $\sim$1035~cm\textsuperscript{$-1$} are
associated with O--H/C--H bending, phenolic C--O, and C--O--C
(glycosidic) stretching, respectively~\cite{ref36,ref37}. Notably, the
C--O--C glycosidic stretching band at $\sim$1035~cm\textsuperscript{$-1$}
is markedly stronger for MP than for GJL, which is consistent with the
higher content of glycosylated anthocyanins and polyols in mangosteen
peel~\cite{ref35}.

The simultaneous presence of hydroxyl (--OH) and carbonyl/carboxyl
(C=O, C--O) groups in both extracts is significant for DSSC operation,
because such polar anchoring groups are commonly associated with
coordination to the TiO\textsubscript{2} surface and with electron
injection from the excited dye into the conduction
band~\cite{ref36}. We note, however, that FTIR identifies the functional
groups but does not by itself quantify the anchoring strength or the
electron-injection efficiency; these are inferred here and would require
electrochemical measurements for direct confirmation. Importantly, the
band positions of MP and GJL coincide without appreciable shifts,
supporting the UV-Vis observation that mixing the two dyes is
predominantly a physical, complementary combination rather than a
chemical reaction that would alter their intrinsic functional groups.

\subsection{Surface and Cross-Sectional Analysis}
\label{sec:morph}

Figure~\ref{fig:3} displays optical micrographs illustrating the
surface morphology of ITO/TiO\textsubscript{2}/dye photoanodes prepared
with varying mixing ratios of MP to GJL natural dyes. Significant
morphological variations are observed across the different ratios. The
MP-rich compositions (3:1 and 2:1) exhibit pronounced cracking and poor
film continuity, characteristic of excessive interfacial stress during
drying. Such structural defects are generally expected to promote charge
recombination and increase series resistance, which is consistent with
the comparatively low fill factors measured for these compositions
(Table~\ref{tab:1}). In contrast, the balanced 1:1 ratio shows a
homogeneous, crack-free morphology with uniform dye coverage, indicating
favorable interfacial compatibility. The GJL-rich formulations (1:2 and
1:3) remain relatively uniform but display incipient granularity and
mild aggregation at higher GJL concentrations, hinting at possible
saturation effects or reduced stability in film formation. Importantly,
optical microscopy is available for the 1:3 optimum (Figure~\ref{fig:3}e),
which shows a largely continuous, crack-free film; this supports a
qualitative correlation between film uniformity and the higher apparent
efficiency of the GJL-rich cells. A formal quantitative image analysis
(e.g., crack-area fraction) was not performed and is left for future
work.

Figure~\ref{fig:4} presents FESEM images characterizing the surface
morphology and cross-sectional structure of
ITO/TiO\textsubscript{2}/dye photoanodes fabricated using pure MP
(3:0), a balanced mixture of MP:GJL (1:1), and pure GJL (0:3) dye
extracts. These three compositions were selected as the two single-dye
end-members and the balanced mid-point; FESEM of the 1:3 optimum was not
acquired in this study and is recommended for future work (its surface
morphology is, however, documented by optical microscopy in
Figure~\ref{fig:3}e). Surface morphological analysis reveals distinct
structural characteristics. The MP film exhibits a densely packed but
irregular surface with localized agglomeration, whereas the GJL sample
displays a more porous yet heterogeneous morphology with microcracks. In
contrast, the mixed dye (1:1) photoanode demonstrates a uniform and
homogeneous nanoporous structure, suggesting improved interfacial
compatibility between the hybrid natural dyes and the
TiO\textsubscript{2} layer. Cross-sectional images show well-adhered
multilayer structures across all samples, with a film thickness of
approximately 20~\textmu m dominated by the mesoporous
TiO\textsubscript{2} layer. The mixed-dye sample exhibits a coherent,
continuous layer with few visible defects, which is consistent with
improved dye coverage and interconnectivity within the
TiO\textsubscript{2} matrix. Taken together with the optical micrographs,
these observations indicate that balanced and GJL-rich compositions form
more uniform films than the MP-rich ones; we interpret this morphological
trend as one plausible contributor to the higher apparent photocurrent of
those cells, rather than as direct proof of enhanced charge injection.

\begin{figure}[htbp]
\centering
\includegraphics[width=0.58\linewidth]{fig-4-optical.pdf}
\caption{Morphological images of the ITO/TiO\textsubscript{2}/dye taken
by optical microscope for the MP:GJL ratio of (a) 3:1, (b) 2:1, (c)
1:1, (d) 1:2, and (e) 1:3.}
\label{fig:3}
\end{figure}

\begin{figure}[htbp]
\centering
\includegraphics[width=0.95\linewidth]{fig-5-fesem.pdf}
\caption{(a--c) Morphological and (d--f) cross-sectional FESEM images of
ITO/TiO\textsubscript{2}/dye for the MP:GJL ratio of 3:0, 1:1, and
0:3.}
\label{fig:4}
\end{figure}

\subsection{Performance of DSSC}

\begin{figure}[htbp]
\centering
\includegraphics[width=0.95\linewidth]{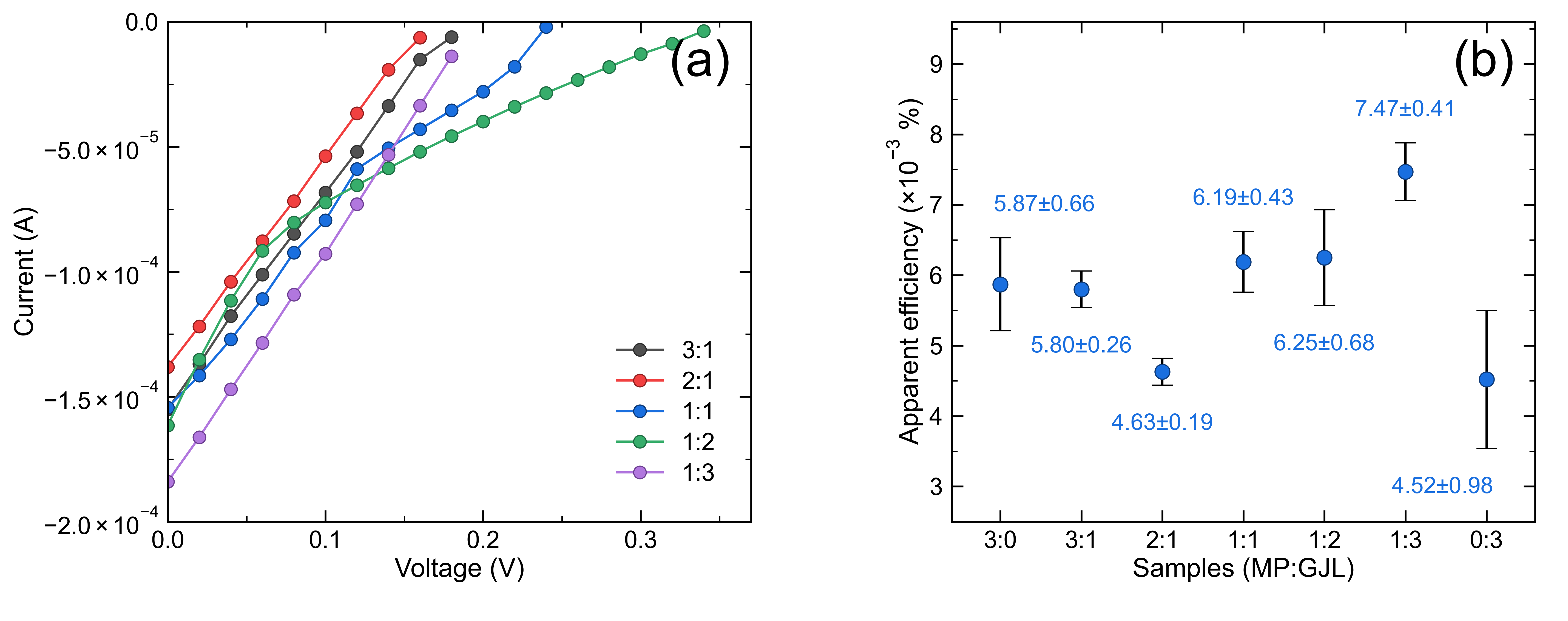}
\caption{$J$--$V$ curves and calculated apparent power conversion
efficiency for DSSC with various mixed dyes of MP:GJL, as well as the
sample controls (3:0 and 0:3).}
\label{fig:5}
\end{figure}

Figure~\ref{fig:5} shows the $J$--$V$ characteristics
and the corresponding apparent power conversion efficiencies (PCE) of
DSSCs sensitized with mixed dyes of MP and GJL at different volume
ratios. The $J$--$V$ characteristics also provide insight into the
working principles of the DSSCs, since several photovoltaic parameters
can be extracted from these curves. The short-circuit current density
($J_{sc}$) reflects the number of photogenerated electrons injected from
the excited dye molecules into the TiO\textsubscript{2} conduction band.
Variations in $J_{sc}$ among the samples indicate that dye composition
affects light harvesting and charge collection. Meanwhile, the
open-circuit voltage ($V_{oc}$) corresponds to the potential difference
between the quasi-Fermi level of electrons in TiO\textsubscript{2} and
the redox potential of the electrolyte, which can be influenced by the
dye/electrolyte interface and charge recombination dynamics. In
addition, the slope of the $J$--$V$ curves near $V_{oc}$ is related to
series resistance and recombination losses, affecting the fill factor
(FF). Therefore, the overall apparent PCE is determined by the interplay
of $J_{sc}$, $V_{oc}$, and FF (see Table~\ref{tab:1}). In this work, the
higher apparent efficiency observed at the 1:3 MP:GJL ratio is
accompanied by the highest $I_{sc}$ (0.18~mA) together with a moderate
$V_{oc}$ and FF, which we interpret as being consistent with improved
photon harvesting; a direct attribution to enhanced electron injection
or suppressed recombination would require electrochemical impedance
spectroscopy (EIS) or incident-photon-to-current efficiency (IPCE)
measurements, which were beyond the scope of this study.

As seen in Figure~\ref{fig:5}a, the photocurrent response varies with
the MP:GJL composition, indicating that the balance between the two
natural dyes plays an essential role in charge generation and transport.
The variation in photocurrent directly influences the overall apparent
photovoltaic performance, which is summarized in Figure~\ref{fig:5}b and
Table~\ref{tab:1}. Averaged over $n=3$ independently fabricated cells,
the apparent efficiency values range from
$(4.52\pm0.98)$ to $(7.47\pm0.41)\times 10^{-3}\%$, with the highest
apparent PCE obtained at the 1:3 MP:GJL ratio
($(7.47\pm0.41)\times 10^{-3}\%$). Because the mean of the 1:3 sample
minus its SD ($7.06\times10^{-3}\%$) exceeds the mean-plus-SD of the
next-best compositions (1:1, $(6.19\pm0.43)\times10^{-3}\%$, and 1:2,
$(6.25\pm0.68)\times10^{-3}\%$, i.e., up to
$\sim6.93\times10^{-3}\%$), the 1:3 composition can be regarded as the
highest-performing among those tested, although the differences among
the intermediate compositions themselves are within roughly two standard
deviations and should be interpreted with caution. This enhancement is
consistent with an effective combination of the two natural dyes, in
which the complementary light absorption and more uniform film formation
plausibly contribute to improved charge generation and collection.

\begin{table}[htbp]
\centering
\caption{Photovoltaic parameters extracted from $J$--$V$ curves of DSSC
with various mixed dyes of MP:GJL. Values are the mean of $n=3$
independently fabricated cells; $\eta$ is reported as mean~$\pm$~SD.
$J_{sc}$ is derived as $I_{sc}/A$ with $A=1.2$~cm\textsuperscript{2}.
Efficiencies are apparent (relative) values measured under
non-standard white-light illumination.}
\label{tab:1}
\begin{tabular}{c c c c c c}
\toprule
Samples (MP:GJL) & $V_{oc}$ (V) & $I_{sc}$ (mA) &
$J_{sc}$ (mA\,cm\textsuperscript{$-2$}) & \textit{FF} (\%) &
$\eta$ ($\times 10^{-3}$ \%) \\
\midrule
3:0 & 0.27 & 0.10 & 0.083 & 28 & $5.87 \pm 0.66$ \\
3:1 & 0.18 & 0.15 & 0.125 & 25 & $5.80 \pm 0.26$ \\
2:1 & 0.16 & 0.14 & 0.117 & 25 & $4.63 \pm 0.19$ \\
1:1 & 0.24 & 0.15 & 0.125 & 21 & $6.19 \pm 0.43$ \\
1:2 & 0.36 & 0.16 & 0.133 & 14 & $6.25 \pm 0.68$ \\
1:3 & 0.18 & 0.18 & 0.150 & 28 & $7.47 \pm 0.41$ \\
0:3 & 0.22 & 0.08 & 0.067 & 32 & $4.52 \pm 0.98$ \\
\bottomrule
\end{tabular}
\end{table}

On the contrary, using single dyes (3:0 or 0:3 ratios) results in
comparatively lower apparent efficiencies ($5.87 \times 10^{-3}\%$ and
$4.52 \times 10^{-3}\%$, respectively). This suggests that while each
dye can sensitize the DSSC individually, the incomplete coverage of the
solar spectrum and possible dye aggregation may limit their performance.
Intermediate ratios such as 1:1 show moderate apparent efficiencies
($6.19 \times 10^{-3}\%$), highlighting the composition-dependent
dye--dye interaction. The lower apparent efficiency at 0:3
($4.52 \times 10^{-3}\%$) may indicate that an excess of GJL does not
favor efficient charge transfer, possibly due to a weaker binding
affinity of GJL molecules compared with MP, leading to reduced dye
anchoring stability on the TiO\textsubscript{2} surface; we present this
as a tentative interpretation rather than a measured conclusion.
Overall, these results indicate that combining MP and GJL dyes,
particularly at the 1:3 ratio, improves the apparent light harvesting
and charge collection relative to single-dye systems under the present
measurement conditions.

To place these values in context, natural-dye DSSCs typically report
power conversion efficiencies in the range of about 0.1--2\% under
calibrated AM1.5G illumination, with co-sensitization strategies
generally yielding the higher end of this
range~\cite{ref17,ref31,ref38,ref39,ref40}. The apparent efficiencies
obtained here are between two and three orders of magnitude smaller.
Although the estimated irradiance of the white-LED bench
($\approx 98$~mW\,cm\textsuperscript{$-2$}) happens to be close to the
AM1.5G reference intensity, its spectrum is not, so these values remain
not directly comparable with AM1.5G-calibrated literature results and
should be read only as internal, relative comparisons; in particular,
the absence of a near-infrared component penalises the chlorophyll-rich
compositions. Several further factors plausibly limit
the absolute performance of the present devices: (i) the use of a
graphite/carbon counter electrode, which typically has a higher
charge-transfer resistance than platinum~\cite{ref20,ref36}; (ii) the
absence of a compact TiO\textsubscript{2} blocking layer, which can
allow back-recombination at the ITO/electrolyte interface; (iii) the
relatively thick, spin-coated TiO\textsubscript{2} film and the limited,
unquantified dye loading from crude extracts; (iv) the KI/I\textsubscript{2}
in ethylene-glycol electrolyte and the open, unsealed cell geometry,
which constrain charge transport and long-term stability; and (v)
non-optimized extraction that may reduce the amount of intact
chromophore~\cite{ref34,ref36}. Addressing these factors, together with
AM1.5G-calibrated measurement, is expected to raise both the absolute
and the relative efficiencies.

\begin{figure}[htbp]
\centering
\includegraphics[width=0.58\linewidth]{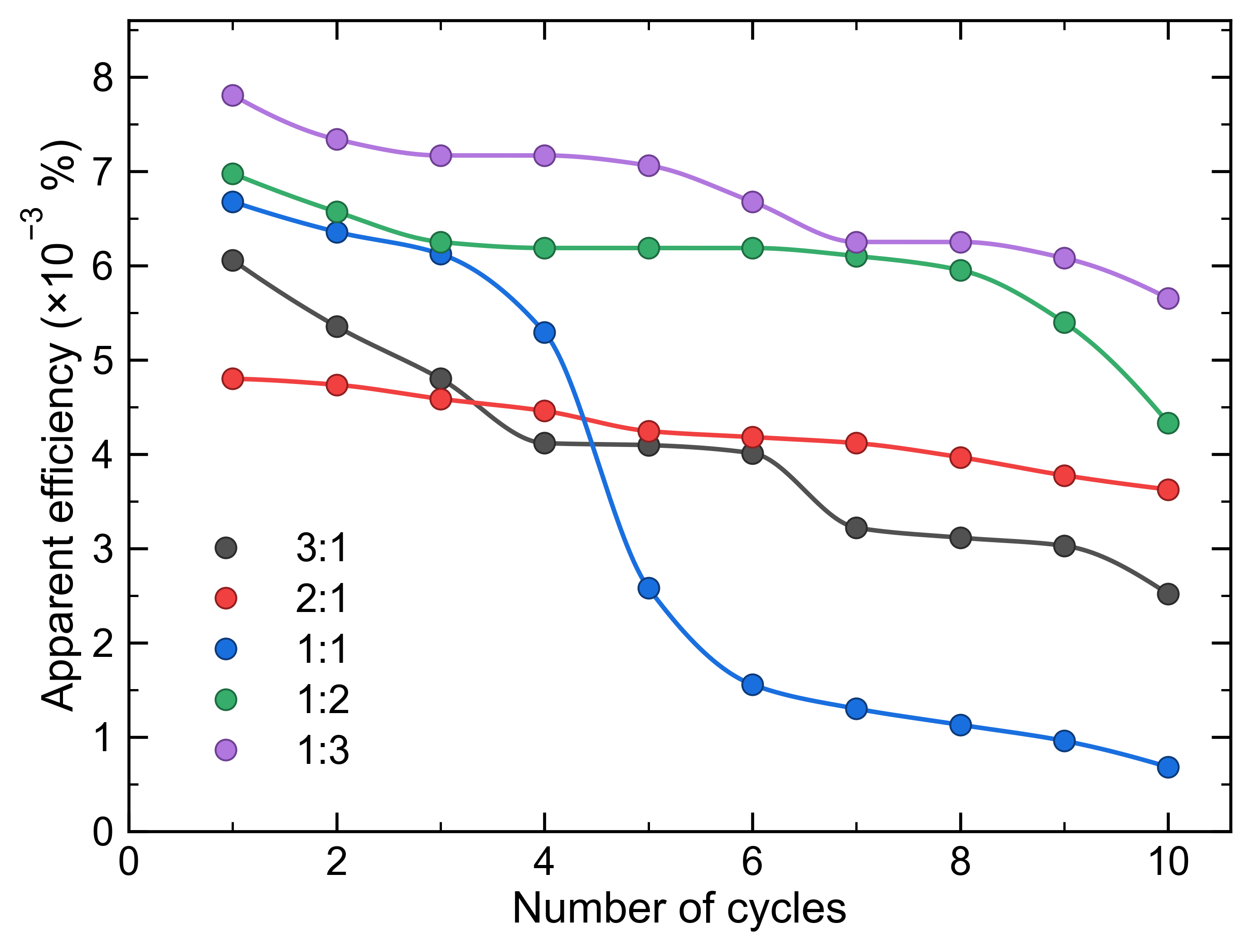}
\caption{Calculated apparent power conversion efficiencies of all
samples over 10 consecutive measurement cycles. Each cycle is one
$J$--$V$ scan recorded at a 10~min interval under continuous white-LED
illumination, corresponding to a total illumination time of
approximately 100~min per cell.}
\label{fig:6}
\end{figure}

Figure~\ref{fig:6} shows the evolution of apparent power conversion
efficiency over ten consecutive measurement cycles for DSSCs
sensitized with mixed MP and GJL dyes at five different volume ratios.
Here, one ``cycle'' denotes a single $J$--$V$ scan of the same cell
recorded at a 10~min interval under continuous white-LED illumination,
so that the ten cycles correspond to a total illumination time of
approximately 100~min without a deliberate dark interval between scans.
Measurements were performed at room temperature, without active
temperature control or active cooling of the cell; the ten cycles
therefore probe the
short-term repeatability and initial photostability of the cells rather
than long-term (hours-to-days) operational stability, which was not
assessed in this study. Two experimentally relevant observations emerge.
First, the samples with a GJL-rich composition (1:3 and 1:2 MP:GJL)
begin with the highest efficiencies and retain a larger fraction of
their initial performance across cycles, whereas the 1:1 mixture
exhibits a rapid decrease in PCE after only a few cycles. Second,
MP-dominant samples (3:1 and 2:1) show intermediate behavior: moderate
initial efficiencies and a steady, less abrupt decay with cycling. These
trends indicate a trade-off between instantaneous performance and
short-term operational stability that depends on dye composition.

A mechanistic interpretation consistent with the observed trends
involves several interrelated processes, which we present as plausible
contributors rather than as directly verified mechanisms: (1)
photobleaching and molecular degradation~\cite{ref32}, which reduce
light absorption and thus photocurrent over repeated illumination; (2)
dye desorption from the TiO\textsubscript{2} surface~\cite{ref33}, in
which weaker binding promotes progressive desorption into the
electrolyte and loss of active dye coverage; and (3) electrolyte--dye
chemistry and regeneration kinetics, where an imbalance between electron
injection and dye regeneration can increase recombination losses during
cycling~\cite{ref34}. These processes would explain why some mixtures
(e.g., 1:3) deliver higher initial apparent PCE and better retention
(suggesting relatively photostable chromophores and/or stronger surface
anchoring), whereas others (notably 1:1) are prone to more rapid
decline. Distinguishing these contributions quantitatively would require
dedicated measurements (e.g., time-resolved absorbance, dye-desorption
assays, and EIS) that are beyond the present scope. Overall, optimizing a
natural-dye DSSC requires balancing spectral coverage and initial
apparent PCE with photochemical and interfacial stability; the
compositional window that maximizes one property may not maximize the
other.

\section{Conclusions}

This study demonstrates that blending natural dyes extracted from
mangosteen peel and grass jelly leaves can improve the
apparent photovoltaic performance of DSSCs.
Among the investigated compositions, the MP:GJL ratio of 1:3 exhibited
the highest apparent power conversion efficiency of
$(7.47 \pm 0.41) \times 10^{-3}\%$ (mean $\pm$ SD, $n=3$), which is
higher than that obtained using individual dyes (MP:
$5.87 \times 10^{-3}\%$ and GJL: $4.52 \times 10^{-3}\%$). The improved
performance is attributed to the complementary light absorption of the
two pigments, which broadens photon harvesting across the visible
spectrum, together with the more uniform, crack-free films formed by the
balanced and GJL-rich compositions and the hydroxyl and carbonyl
anchoring groups identified by FTIR. These results indicate that
combining natural dyes derived from abundant plant resources can be a
practical strategy for developing low-cost and environmentally friendly
DSSC sensitizers, offering a sustainable pathway for utilizing
agricultural by-products in solar energy technologies. It should be
emphasized that the efficiencies reported here are apparent (relative)
values obtained under non-standard, luxmeter-estimated illumination and
from three cells per composition; they are intended for internal
comparison rather than as absolute figures of merit. Future work should
therefore include AM1.5G-calibrated measurements, quantification of
chromophore concentration and dye loading, electrochemical analysis
(EIS/IPCE) to substantiate the charge-transfer mechanism, FESEM of the
1:3 optimum, and long-term stability testing, in order to improve dye
stability and adsorption on semiconductor surfaces and to enhance the
long-term performance of these bio-sensitized solar cells.

\section*{Data Availability Statement}

The datasets generated during and/or analyzed during the current study
are available from the corresponding author.

\section*{Declaration of Competing Interest}

The authors declare that they have no known competing financial
interests or personal relationships that could have appeared to
influence the work reported in this paper.

\section*{Declaration of Generative AI and AI-Assisted Technologies in the Writing Process}

During the preparation of this work, the authors used Claude
(Anthropic), a large language model, accessed through the Claude
Code application (model Claude Opus 5.5 for the revision stage).
It was used (i) to improve the English language and readability of the manuscript text; and (ii) to assist with typesetting and the formatting of the figures. The authors chose to use this tool because the authors wished to improve the clarity and correctness of the English writing. The tool was not used to generate, analyze, or interpret the experimental data, nor to create or alter the image content of any figure. After using this tool, the authors reviewed and edited the content as needed and take full responsibility for the content of the published article.

\ifblind\else
\section*{Author Contributions}

All authors contributed to the submitted manuscript. \textbf{EN}
contributes to conceptualization, methodology design, resources,
writing of original drafts, supervision, and funding acquisition.
\textbf{HS} and \textbf{DIB} contribute to conducting research and
visualization. \textbf{WSS} contributes to the supervision and review
of the original draft. \textbf{MAG} and \textbf{RA} contribute to the
funding acquisition. \textbf{TT} and \textbf{AR} contribute to
resources. \textbf{MY} and \textbf{RK} contribute to resource and
funding acquisition.

\section*{Acknowledgement}

EN thanks Institut Teknologi Sumatera (Itera) for supporting this
research via ``Riset Kolaborasi Indonesia (RKI) 2025'' scheme with
contract no. 1998by/IT9.2.1/PT.01.03/2025.
\fi


\end{document}